\documentclass[superscriptaddress,showpacs,preprintnumbers,amsmath,amssymb,
nofootinbib,twocolumn,aps,prd,10pt]{revtex4-1}
\usepackage{amssymb,amsmath}
\usepackage{epsfig}
\usepackage{xcolor}
\usepackage[utf8]{inputenc}
\usepackage{stmaryrd}
\usepackage{mathrsfs}
\usepackage{mathalfa}
\usepackage[normalem]{ulem}

\newcommand{\ve}{\varepsilon}

\newcommand{\be}{\begin{equation}}
\newcommand{\ee}{\end{equation}}
\newcommand{\ba}{\begin{eqnarray}}
\newcommand{\ea}{\end{eqnarray}}
\newcommand{\beg}{\begin{gather*}}
\newcommand{\eng}{\end{gather*}}
\newcommand{\hh}{,\hspace{0.5cm}}
\newcommand{\hhh}{,\hspace{0.2cm}}

\newcommand{\n}[1]{\label{#1}}

\newcommand{\he}{\hat{\eta}}

\newcommand{\hm}{\hat{m}}

\newcommand{\hz}{\hat{z}}
\newcommand{\hq}{\hat{q}}
\newcommand{\hF}{\hat{\cal F}}
\newcommand{\hFF}{\hat{F}}
\newcommand{\hV}{\hat{V}}
\newcommand{\cF}{{\cal F}}

\def\XXint#1#2#3{{\setbox0=\hbox{$#1{#2#3}{\int}$ }
\vcenter{\hbox{$#2#3$ }}\kern-.6\wd0}}

\usepackage[makeroom]{cancel}
\usepackage[caption=false]{subfig}
\usepackage[colorlinks=true,
            citecolor=green,
            linkcolor=red,
            filecolor=cyan,
            urlcolor=magenta,
            backref=false]{hyperref}

\begin{document}

\title{Nonlocal scalar field in an external potential: WKB approximation }

\author{Valeri P. Frolov}
\email{vfrolov@ualberta.ca}
\affiliation{Theoretical Physics Institute, University of Alberta, Edmonton, Alberta, Canada T6G 2E1}


\begin{abstract}
We consider a nonlocal theory of a scalar massive field in a flat spacetime background in the presence of an external potential and construct WKB solutions for this theory. We use a model in which the kinetic part of the scalar field action is modified by changing $\Box$ to $\Box f(\Box)$ operator. We discuss conditions when the corresponding form factor $f$ is chosen so that the theory does not contain new unphysical degrees of freedom.
We applied the obtained WKB solutions for study  energy levels of the field trapped by a one-dimensional potential and the probability of the barrier penetration. This allows us to illustrate how the effects of the nonlocality change the known results obtained for the local field theory.
\end{abstract}


\maketitle

\section{Introduction}

The idea of nonlocality is quite old in the theoretical physics. Nonlocal modifications of the field theory were discussed already in the publications \cite{Wataghin:1934ann,Yukawa:1950eq,Yukawa:1950er}.
Even if one works with a local quantum field theory, already at the one loop level its effective action is nonlocal. This nonlocality is connected with the  vacuum polarization and particle creation effects and it reflects a fact that a  vacuum is a physical medium and in this sense it behaves as a condensed matter with very special properties. A nonlocality that we consider in this paper is of a different origin. We assume that a field theory is described by a nonlocal action already at the tree level. In particular this means that its interaction with external sources as well as a self-interaction is nonlocal.

Usually such a nonlocal modification contains a scale parameter   which determines either the energy at which the effects of the nonlocality become important or the corresponding spacetime length when it happens.
Nonlocal theories were discussed in the beginning of seventies of the past century by Efimov \cite{Efimov:19,Efimov:1972wj,Efimov:18,Alebastrov:1973np}.
The nonlocality is often introduced to modify  high energy properties of the theory and  to improve its ultraviolet behavior. More recently the interest to nonlocal theories increased. This was mainly stimulated by the development of the string theory.
Nonlocal fields naturally arise in the string theory and in the theories with a noncommutive geometry  (see e.g. \cite{Witten:1985cc,Frampton:1988kr,Eliezer:1989cr,Kostelecky:1990mi,Harms:1993bt,Tseytlin:1995uq,Calcagni_2014,Calcagni_2015} and references therein).
Nonlocal effects in quantum gravity were  widely discussed recently. A comprehensive review
of modifications of gravity involving a minimal length scale and related  references can be found  in \cite{Hossenfelder:2012jw}.

There is a subclass of the non-local field theories  sometimes called {\it ghost-free}  theories. An effective action in these models contains an infinite number of derivatives so that the corresponding field equations are effectively nonlocal.
These models have been widely discussed recently and they have rather "nice properties". In these theories the nonlocality is introduced in such a way that it preserves local Lorentz invariance and it does not introduce new unphysical degrees of freedom (ghosts). In these models the ultraviolet (UV) behaviour of the theory at short distances is improved, while in the infrared (IR) regime (at large scales) they reproduce results of a corresponding local theory  \cite{Tomboulis:1997gg,Biswas:2011ar,Modesto:2011kw,Biswas:2013cha,Shapiro:2015uxa,Biswas:2010zk,Biswas:2013kla}.
Main motivation for study such infinite derivative modifications of the gravity equations is connected with attempts to solve  long-standing problems  of  cosmological and black hole singularities \cite{Biswas_2006,Koshelev_2007,Biswas_2010,Modesto_2011,Biswas:2011ar,Conroy:2015nva,
Modesto:2017sdr,Buoninfante:2018xiw,Koshelev:2018hpt,Kilicarslan:2018unm,Koshelev_2019,Kilicarslan:2019njc}.

Complete equations of a modified gravity which include both nonlocality and nonlinearity are quite complicated. Much easier for study is a linearized version of the theory. However already study of linearized models allows one to obtain several stimulating results: (i) The nonlocality removes singularities of the field produced by a point-like sources \cite{Biswas:2011ar,delaCruz-Dombriz:2018aal,Buoninfante:2019swn,Boos:2018bxf,Boos:2020ccj}; (ii) There exists a mass gap for mini black hole formation \cite{Frolov:2015bta,Frolov:2015bia}; (iii) It allows one to demonstrate a formation of the inner horizon for black hole creation in the scattering of ultrarelativistic particles \cite{Frolov:2015usa}.
A standard technique for solving these problems is usage of nonlocal  Green functions which for the flat spacetime background can be found by means of the Fourier transform.

Study of  the nonlocal field in the presence of an external potential is a much more complicated problem.
There exist very special cases when it is possible to find an explicit solution. For example, such a solution was obtained for scattering of a nonlocal scalar field by a delta-like potential \cite{Boos:2018kir,Frolov:2018bak,Buoninfante:2019teo}.
However in a general case in the presence of an arbitrary potential one needs to solve a nonlocal linear equation with the space-dependent coefficients which is a very non-trivial problem.

The purpose of this paper is to demonstrate that one can obtain asymptotic solutions for this problem by using a standard WKB method. Its main idea  is to search for a solution $\Phi(x)$ of the field equation in the form $\Phi(x)\approx u(x) \exp{[iS(x)/\hbar]}$. After substitution of this ansatz into the field equation one collects terms of the expansion into the powers of $\hbar$  which are of the same order of $\hbar$. In the leading order one gets  a first-order partial differential equation of the form $H(\nabla S,x)=0$ known as an eikonal equation. A sub-leading equation determines evolution of a slowly changing field amplitude $u(x)$. The eikonal equation can be identified with the Hamilton-Jacobi equation for
the Hamiltonian obtained by the substitution $\nabla S=p$ into it. Initial data for the Hamilton-Jacobi equation specify a beam of  trajectories in the phase space which forms a Lagrangian submanifold (for details see e.g. a remarkable book \cite{Arnold:1989}). Knowledge of this Lagrangian submanifold allows one not only to construct the eikonal function $S(x)$ but also to find a solution of the transport equation for the amplitude $u(x)$ by using the Liouville theorem. This WKB method is widely used in the standard quantum mechanics and field theory where the corresponding equations are second order partial differential equations. However, it can be applied to a wider class of so called quasilinear differential equations.
A comprehensive presentation of these results can be found in the book \cite{maslov1981semi}. In these paper we apply this method for study quasiclassical solutions of the linear nonlocal scalar field equations in the presence of an external potential. A similar approach for other higher and infinite order equations can be found in \cite{Lv:2017pun}.

This paper is organized as follows. In section~II we describe a model of a nonlocal scalar field with infinite number of derivatives which is  analysed in the paper. In section~III we construct a WKB solution for such a field in the presence of an external potential in any number of spacetime dimensions. In section~IV we consider a special case when the potential depends on only one spatial Cartesian coordinate. Energy levels for the nonlocal scalar field confined by the one-dimensional parabolic potential are calculated in section~V. Under-barrier propagation of the nonlocal field and barrier penetration effect are discussed in section~VI. Section~VII contains discussion of the obtained results. Additional technical details are collected in two appendices.

\section{Nonlocal scalar field equation}

We consider $N$-dimensional flat spacetime. Its metric in Cartesian coordinates  is
\be
ds^2=\eta_{\mu\nu}dx^{\mu} dx^{\nu}\hh  \mu,\nu=0,1,\ldots,N-1\, ,
\ee
where $\eta_{\mu\nu}=\mbox{diag}(-1,1,\ldots,1)$.
Let us consider a scalar massive field $\varphi$ obeying the Klein-Gordon equation
\be\n{KG}
(\hbar^2\Box\  -m^2-V)\varphi =0\, .
\ee
Here $\Box=\eta^{\mu\nu}\partial_{\mu}\partial_{\nu}$,
$m$ is the mass of the field and $V(x)$ is an external potential. In what follows we shall study solutions of this equation and its nonlocal generalization in the WKB approximation. For this reason we keep the Planck constant $\hbar$, while as usual put $c=1$\footnote{In these units one has
$[\hbar]=M L$, $ [\Box]=L^{-2}$, and $ [m^2]=[V]=M^2$ .}.  The Klein-Gordon equation follows from the action
\be\n{WKG}
W[\varphi]={1\over 2}\int d^N x \varphi (F(\Box) -V)\varphi\, ,\
F(\Box)=\hbar^2\Box -m^2 \, .
\ee

We consider a generalization of the Klein-Gordon equation in which the operator $F(\Box)$ is modified. Namely, we assume that it is a scalar operator which may contain an arbitrary (finite or infinite) number of partial derivatives $\partial/\partial x^{\mu}$. It is easy to check that the covariance of the action requires that such an operator can be written as a scalar function of the $\Box$-operator. In order to keep a proper dimensionality of the action we introduce a parameter $\mu$ which has a dimension of the mass and we write the operator $F(\Box)$ in the form
\be \n{FM}
F(\Box)=\mu^2 \hFF(\hbar^2\Box/\mu^2)\, .
\ee

The corresponding generalized Klein-Gordon equation is
\be \n{GFKG}
[-\hFF(z)+\hV]\varphi =0\, ,
\ee
where $z=\hbar^2 \Box/ \mu^2$ and $\hV=V/\mu^2$.

We consider a  class of theories for which the function $\hFF(z)$  has the following properties:
\begin{enumerate}
\item $\hFF(z)$ does not vanish anywhere on the complex plane of $z$ besides the  point $z=\hm^2$;
\item $\hFF|_{ z=\hm^2}=0$ and $d\hFF(z)/dz|_{ z=\hm^2}=1$;
\item For a real value of its argument $z$ the function $\hFF(z)$ is real.
\end{enumerate}
These conditions guarantee that the inverse of the operator $\hFF$ has a single pole with its residue equal to 1.
In other words, there is no new unphysical (ghost) degrees of freedom in this theory.
To satisfy the required properties it is sufficient to choose  $\hFF$ in the form
\be
\hFF(z)=\cF(\hz)\hh \cF(\hz)=\hz f(\hz)\hh \hz=z-\hm^2\, ,
\ee
where $f(\hz)=\exp(g(\hz))$ and $g(\hz)$ is an entire function of the complex variable satisfying the condition
$g(\hz=0)=0$. We call $f(\hz)$ a form factor.

In order to illustrate the results we shall use a special example of the form factor. Namely we put
$g(\hz)=\hz^2$ then one has
\be\n{cfz}
\cF(\hz)=X(\hz)\equiv \hz \exp(\hz^2)\, .
\ee
This function obeys all the conditions listed above. Besides this for real $\hz$ it is a monotonically increasing from $-\infty$ at $\hz=-\infty$ to $+\infty$ at $\hz=+\infty$ and is positive for $\hz>0$. One also has $\cF(-\hz)=-\cF(\hz)$. The function $X(x)$ is shown at figure~\ref{P0}. Other properties of this function which are used later in the paper can be found in appendix~B.

\begin{figure}[hbt]
    \centering
    \vspace{10pt}
    \includegraphics[width=0.48\textwidth]{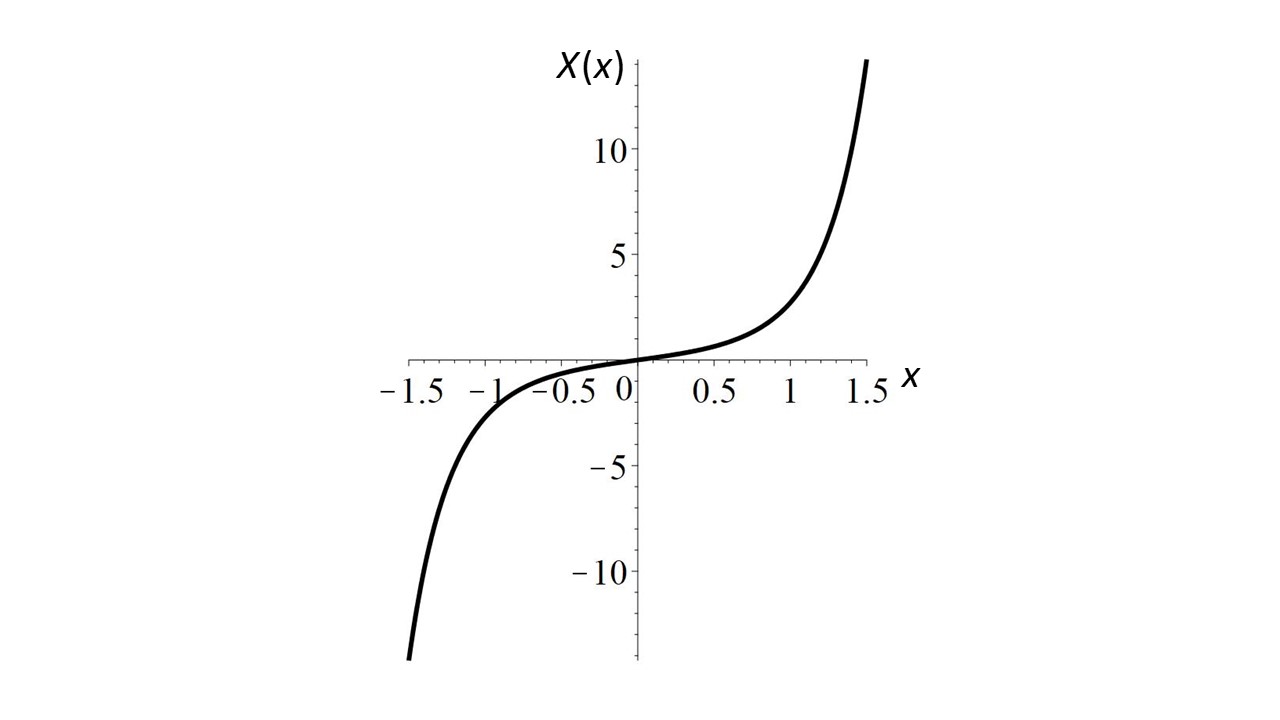}
    \caption{Function $X(x)$.}
    \label{P0}
\end{figure}

One can expand the function $\cF(\hz)$ into powers of  $\hz$. This gives an infinite series of the powers of $\Box$ for the field operator $\hat{F}(\hbar^2\Box/\mu^2)$. This means that this operator is nonlocal. Let us also mention that the mass parameter $\mu$ which enters this operator  determines a characteristic energy scale at which the effects of the nonlocality become important. In the limit $\mu\to  \infty$ the operator $F$ becomes $\hbar^2\Box -m^2$ and the theory is local.

\section{WKB approximation}

\subsection{A solution of the nonlocal field equation in the WKB approximation}

Let us write the nonlocal field equation (\ref{GFKG}) in the form\footnote{\n{mu}
In the nonlocal theory it is convenient to use the  parameter $\mu$ to define
dimensionless quantities.  We specify such quantities by using a hat over them. Here and later we use the following notations
\be
\hm=m/\mu\, , \  \he=\eta/\mu\, , \   \he_0=\eta_0/\mu\, , \hat{V}=V/\mu\, ,\  \hz=\he_0^2-\he^2\, .\nonumber
\ee
}
\be\n{ooo}
\hat{\cal O}\varphi=0\hhh
\hat{\cal O}={1\over 2}(-\mu^2\cF(\hz)+\hV)\hhh
\hz=\hbar^2\Box/\mu^2 -\hm^2
\, .
\ee
We include a factor $1/2$ in the operator $\hat{\cal O}$ which evidently does not change solutions but this simplifies the form of some relations. The operator $\hat{\cal O}$  belongs to the class of equations (\ref{OO}) discussed in the appendix.
We are looking for a solution of the field equation (\ref{ooo}) in the WKB approximation and write it in the following form
\be \n{vvv}
\varphi(x)=\exp\left({i S(x)\over \hbar}\right)\sum_{j=0}^{\infty} \left({\hbar\over i}\right)^j  u_j(x)\, .
\ee
Here $S(x)$ is a fast changing phase, while $u_j(x)$ are slowly changing functions. Let us emphasize that besides standard conditions of the validity of the WKB approximation formulated in  textbooks on quantum mechanics we also assume that the parameter of nonlocality $\mu$ does not depend on $\hbar$. We  keep this parameter fixed in the limit $\hbar\to 0$. This allows us to discuss effects of the nonlocality in the quasiclassical approximation.

In order to find the eikonal function $S(x)$ and the amplitudes $ u_j(x)$ we follow the steps described in  appendix~A. First of all we define a symbol $H(x,p)$ of the operator $\hat{\cal O}$. Let  $p_{\mu}$ be a covector  and denote $p^2=p_{\mu}p^{\mu}$.  Then one should substitute $-\hbar^2 \Box\to p^2$ into (\ref{ooo}). The corresponding symbol for the operator $\hat{\cal O}$ is
\be \n{HHH}
H(x,p)={1\over 2}\left[-\mu^2\cF(\hz)+V(x)\right]\hhh \hz=-{1\over \mu^2}(p^2+m^2)
\, .
\ee
The eikonal equation (\ref{HJ}) implies that the following constraint $H(x,p)=0$ is valid.

The next step is to study a dynamical system in $2N$-dimensional phase space satisfying the Hamilton's equations
\be \n{HEE}
{d x^{\mu}\over  d\tau}={\partial H\over \partial p_{\mu}}\hh
{d p_{\mu}\over  d\tau}=-{\partial H\over \partial x^{\mu}}\, .
\ee
Since the Hamiltonian $H(x,p)$ is an integral of motion it is sufficient to choose $H(x,p)=0$ at the initial "time" $\tau=0$, then the constraint $H(x,p)=0$  is valid for any later "time".

We define a $(N-1)$-dimensional surface $\Sigma$ by conditions
\be
x^{\mu}=f^{\mu}(y^i)\hh i=1,\ldots,N-1
\, ,
\ee
and choose an initial state for the field $\varphi$ in the form
\be
\left. \varphi(x)\right|_{\Sigma}=u_0(y)\exp \left[{ i\over \hbar} S^0(y)\right]\, .
\ee
After solving equations (\ref{ICS}) we impose the initial conditions (\ref{IC}) for the Hamiltonian equations (\ref{HEE}). Solving this system we find
\be
x^{\mu}=x^{\mu}(\tau,y^i)\hh p_{\mu}=p_{\mu}(\tau,y^i)\, .
\ee
A sought solution $\varphi(x)$ of the nonlocal equation (\ref{ooo}) in the leading order of the WKB approximation is
\be \n{QCC}
\varphi(x)=u_0(y) \sqrt{ J(0,y)\over J(\tau,y)} \exp \left[{ i\over \hbar}\left( S^0(y)+\int_{0}^{\tau} p_{\mu} dx^{\mu}\right)\right] \, .
\ee
This expression is obtained from (\ref{QC}) by using the relation
\be
 {\partial^2 H(x,p)\over \partial x^{\mu}\partial p_{\mu} }=0
\ee
valid for the Hamiltonian (\ref{HHH}). The integral in the exponent in (\ref{QCC}) is taken over phase trajectories which belong to the Lagrangian submanifold determied by given initial conditions.
Here the function $J$ which enters into the prefactor in (\ref{QCC}) is defined by (\ref{JJJ}).

\subsection{Remarks on the Hamilton's equations}

Let us discuss now the Hamilton's equations in more detail. One has
\be
{\partial H\over \partial p_{\mu}}= \cF' p^{\mu}\, .
\ee
Here a prime denotes a derivative of $\cF(\hz)$ with respect to its argument $\hz$.
Hence the Hamilton's equations (\ref{HEE}) take the form
\ba \n{HAM1}
&&\dot{x}^{\mu}\equiv {dx^{\mu}\over d\tau}= \cF' p^{\mu}\, ,\\
&&\dot{p}_{\mu}\equiv{dp_{\mu}\over d\tau}=-{1\over 2}V_{,\mu}\, .\n{HAM2}
\ea

The first of these equations implies
\be \n{xz}
\dot{x}^2\equiv \eta_{\mu\nu}\dot{x}^{\mu}\dot{x}^{\nu}= (\cF ')^2 p^2\, .
\ee
Since $\hz=-(p^2+m^2)/\mu^2$ the right-hand side of this relation is a function of $p^2$.
We assume that one can solve  equation (\ref{xz}) and find $p^2$ as a function of $\dot{x}^2$.
Then the Lagrangian of this dynamical system is
\ba \n{LL}
&&L(x,\dot{x})=\dot{x}^{\mu}p_{\mu}-H\nonumber\\
&&=
{1\over2}[\mu^2(\cF-2\hz\cF')-2m^2\cF']-{1\over2}V(x)\, .
\ea
It is understood that in this expression for $L$ one should express $\hz$ as a function of $\dot{x}^2$.
Since $L$ does not contain explicitly the parameter $\tau$  the corresponding "energy" for this Lagrangian is conserved. This "energy" is nothing but the Hamiltonian in which $p^2$ is a function of $\dot{x}^2$ defined by (\ref{xz}).
The constraint equation in the dimensionless form is (see footnote~\ref{mu}) )
\be\n{FV}
-\cF(\hz)+\hat{V}(x)=0\, .
\ee
If $\hV(x)\ge 0$ then $\cF(\hz)=\hz f(\hz)$ should be real and positive. The argument of this function $\hz=-(p^2+m^2)/\mu^2$ is real. Since $f(0)=1$ and $f(\hz)$ does not vanish this function is positive for real $\hz$. This means that the sign of $\cF(\hz)$ coincides with the sign of $\hz$. The equation (\ref{FV}) implies that the classical motion is possible only in a domain where $\hz\ge 0$.

In the absence of the potential the constraint equation takes the form
\be
H_0(p)=-{1\over 2}\mu^2 \hz f(\hz)=0 \, .
\ee
Since $f(\hz)$ does not vanish this equation  gives $p^2=-m^2$. The condition $\mu^2\cF'(\hz=0)=1$ implies that
\be
\left. {\partial H_0(p)\over \partial p_{\mu}}\right|_{p^2+m^2=0}=p^{\mu}.
\ee
This means that in the absence of the potential the Hamilton's equations (\ref{HAM1})-(\ref{HAM2}) coincide with the equations for a free particle with the Hamiltonian $(1/2)(p^2+m^2)$. In other words, the effects of the nonlocality are important only off-shell, that is in the presence on the external potential.

\section{One-dimensional case}

To illustrate an application of the general approach described in the previous section we consider now  a simple  model. Namely, we assume that the external potential $V(x)$ depends  on only one spatial Cartesian coordinate. We denote this coordinate by $q$ and use the following notations
\be
x^{\mu}=(t,q,x_{\perp})\hh x_{\perp}=(x^2,\ldots,x^N)\, .
\ee
We use index $ a=2,\ldots,N$ to enumerate transverse coordinates $x_{\perp}$.
We write the eikonal function $S$ which enters the WKB solution (\ref{vvv}) in the form
\be\n{1d}
S= -\ve t+(p_{\perp},x_{\perp}) +{\cal S}(q)\hhh (p_{\perp},x_{\perp})=\sum_{a=2}^N p_a x^a  \,  .
\ee
We also assume that the functions $u_j$ depend only on the coordinate $q$.

Let us assume first that the potential $V(q)$ vanishes. In order to solve the corresponding field equation (\ref{ooo})
one can put ${\cal S}(q)=\eta q$ where $\eta=$const,  $u_0$=const and $u_{j\ge1}=0$. This is a plane wave solution of the field equation (\ref{ooo}).
Denote $p_{\mu}=(-\ve,\eta,p_{\perp})$, where the components of the covector $p_{\perp}$ are $p_a$ with
$a=2,\ldots,N$. Hence $S=(p,x)=p_{\mu}x^{\mu}$. Other solutions of the field equation  can be obtained by superimposing the plane-wave solutions, which in fact is just by using the Fourier representation. The constraint equation $p^2+m^2=0$ gives
\be
-\ve^2+p_{\perp}^2+m^2+\eta^2=0\, .
\ee

In the presence of the potential we look for a solution of the nonlocal field equation (\ref{ooo}) in the form
\be
\varphi(x)=\exp\left[ {i\over \hbar}\left(-\ve t+(p_{\perp},x_{\perp})\right) \right] \Phi(q)\, .
\ee
Substituting this expression  into  (\ref{ooo})  one obtains the following reduced equation
\ba
&&\hat{\cal O}_1 \Phi=0\, ,\n{O1}\\
&&\hat{\cal O}_1={1\over 2}[ -\mu^2\hat{D}  f(\hat{D})+V(q) ] \, , \\
&&\hat{D}={1\over \mu^2}\left[\eta_0^2 +\hbar^2{\partial^2\over \partial q^2}\right]\hhh
\eta_0^2=\ve^2-p_{\perp}^2-m^2
\, .
\ea

To obtain the symbol $H_1(q,\eta)$ of the operator $\hat{\cal O}_1$ we substitute
\be
{\hbar\over i}{\partial \over \partial q}\to \eta\, .
\ee
The result is
\ba  \n{H1}
&&H_1(q,\eta)={1\over 2}\left[ -\mu^2 {\cal F}(\hz)+V(q)\right]\, ,\\
&&{\cal F}(\hz)=\hz f(\hz)\hhh \hz=\he_0^2-\he^2\, .
\ea

For real $\eta$ the variable $\hz$ is also real as well as the Hamiltonian itself. We also require that for real $\hz$ one has $\cF'(\hz)> 0$. For $f(\hz)=\exp(g(\hz))$ this condition is satisfied when
\be
1+\hz {dg(\hz)\over d\hz}>0.
\ee
 In particular, this is valid for the choice (\ref{cfz}).

The Hamilton's (\ref{HAM1})-(\ref{HAM2})  in $(q,\eta)$-phase space reduce to
\be  \n{HJR}
\dot{q}={\partial H_1\over \partial \eta}=\mu^2\cF' \eta\hhh
\dot{\eta}=-{\partial H_1\over \partial q}=-{dV\over dq}\, ,
\ee
while the constraint equation is $H_1(q,\eta)=0$. Let us denote by $(q(t),\eta(t))$ a solution of these equations with the initial condition $q(0)=q_0$ and $\eta(0)=\eta_0$. A solution of the field equation (\ref{O1}) in the leading order of the WKB approximation is
\ba  \n{QC1}
\Phi(q)&=&{1\over \sqrt{J}} \exp[ {i\over \hbar}{\cal S}(q)] \Phi_0(q_0)\, ,\\
{\cal S}(q)&= &\int_{q_0}^{q} \eta dq=\int_{0}^{\tau} \eta(\tau) \dot{q}(\tau) d\tau \, .
\ea
In the expression for ${\cal S}(q)$ the integral is taken along the path $q(\tau)$ connecting the initial point $q_0$ with a final point $q$. A function $\Phi_0(q_0)=\exp[ {i\over \hbar}{\cal S}(q_0)] u_0(q_0)$ is the initial value for $\Phi(q)$. The factor $J$ is
\be \n{JJJJ}
J=\left| {dx\over d\tau}\right|=\mu \cF' \sqrt{\hz_0-\hz}\hh \hz_0=\he_0^2\, .
\ee
The second equality in the expression for $J$ is obtained by using the first Hamilton's equation (\ref{HJR}).

Since the Hamiltonian $H_1(q,\eta)$ is invariant under the reflection $\eta\to -\eta$ one can  write another WKB solution by a simple change of the sign in the exponent of (\ref{QC1}). Taking a superposition of these two WKB solutions one can write $\Phi(q)$ in the following form
\be  \n{PH}
\Phi(q)={C_1\over \sqrt{J}} \exp[ {i\over \hbar} {\cal S}(q)]
+{C_2\over \sqrt{J}} \exp[ -{i\over \hbar} {\cal S}(q)]\, .
\ee
For $C_2=\bar{C}_1$ the WKB solution (\ref{PH}) is real.

\section{Bound motion: Energy levels}

\subsection{A general case}

Let us assume that the potential $V(q)$ is nonnegative and at some point $a$ the following condition is valid
\be\n{BM}
\hF(\he_0)=\hV(a)\, .
\ee
Then the constraint equation
\be \n{CE}
\hF(\hz)=\hV(q)\,
\ee
implies that the momentum $\eta$ vanishes at $a$. This means that $a$ is a turning point. If $dV/dq (a)>0$ then the "particle" comes to this point from the left, that is from the domain where $q<a$ and $\eta>0$. After reaching this turning point $\eta$ changes its sign and the "particle" moves to the left. If the potential $V(q)$ has another turning point $b$ where  $dV/dq (b)<0$, then the "particle" comes to it from the right with $\eta<0$ changes its direction of motion at $b$ to the opposite and moves to the right with $\eta>0$.

\begin{figure}[hbt]
    \centering
    \vspace{10pt}
    \includegraphics[width=0.48\textwidth]{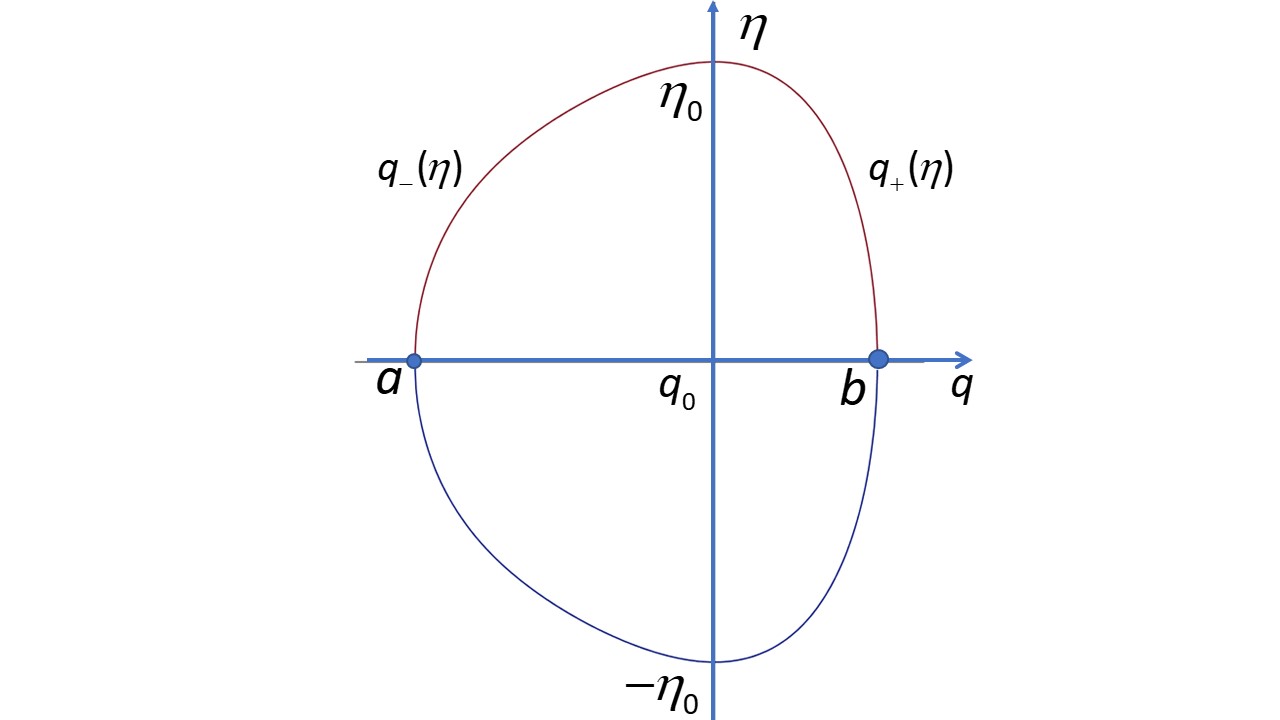}
    \caption{A phase trajectory}
    \label{P1}
\end{figure}

Let us assume that there exist two turning points $a<b$ such that the motion is restricted by the interval $a\le q\le b$. Such a motion is called bound. Let us consider a nonnegative potential $V(q)$ which has its minimum at $q_0$ and $V(q_0)=0$. At this point the absolute value of the momentum $|\eta(q)|$ reaches its maximum which is equal to $\eta_0$. Thus for such a bound motion the phase trajectory lies within a rectangle $[a,b]\times [-\eta_0,\eta_0]$. Let us emphasize that the Hamiltonian (\ref{H1}) contains a free parameter $\eta_0$ which determines a size and a shape of the phase trajectory. Since the parameter $\hz$ depends on the square of the momentum $\eta$, a phase trajectory on the plane $(q,\eta)$ is symmetric with respect to a reflection $\eta\to -\eta$. For a given value $\eta_0>0$  constraint equation (\ref{CE}) defines a closed curve on $(q,\eta)$ phase plane (see figure~\ref{P1}).

The energy levels of the  states corresponding to the bound motion are quantized. In the quasiclassical approximation these levels can be found by using standard Bohr-Sommerfeld quantization condition \cite{maslov1981semi}
\be\n{BS}
J=\oint \eta dq=\pi\hbar (2n+1)\,
\ee
Here the integral is taken over a complete period. For a given $\eta_0>0$ this integral is equal to the surface area inside the phase-trajectory curve  corresponding to this parameter. Because of the reflection symmetry $\eta\to -\eta$ this integral can be written as
\be \n{int}
J=2\int_{a}^{b} \eta(q) dq\, .
\ee
The integral is taken between the turning points $a$ and $b$.

Sometimes it is convenient to rewrite $J$ in another form. Using constraint equation (\ref{CE}) one can express the coordinate $q$ as a function of $\eta$. This function has two branches which we denote by $q_{\pm}(\eta)$  (see figure~\ref{P1}). Then, integrating by parts (\ref{int}) and taking into account that $\eta(a)=\eta(b)=0$ one finds
\be \n{Ja}
J=2\int_{0}^{\eta_0} (q_+(\eta)- q_-(\eta))d\eta \, .
\ee
If the potential $V(x)$ is a symmetric function of $q$, $V(-q)=V(q)$, then one has $q_0=0$ and $q_-(\eta) =-q_+(\eta)$. In this case the quantity $J$ can be written in the form
\be\n{JS}
J=4\int_{0}^{\eta_0} q_+(\eta) d\eta \, .
\ee

\subsection{Parabolic potential: Local theory}

Let us consider a case when the potential $V(q)$ is parabolic
\be \n{VC}
V(q)=V_1^2 q^2\, ,
\ee
where $V_1$ is a positive constant which has the dimension $[V_1]=M/L$.
For the local theory the form factor $f(\hz)=1$ and the constraint equation takes the form
\be \n{ceq}
\eta^2+V_1^2 q^2=\eta_0^2\, .
\ee
Solving (\ref{ceq}) we find
\be
q=\pm V_1^{-1} \sqrt{ \eta_0^2- \eta^2}\, .
\ee
The integral (\ref{JS}) can be easily calculated. We denote the result by $J_0$. Then one has
\be\n{JL}
J_0={\pi\  \eta_0^2\over V_1}\, .
\ee
This gives the following expression for the energy levels
\be
\ve_n^2=m^2+p_{\perp}^2 +\hbar V_1 (2n+1)\, .
\ee
It is easy to see that the number of levels $\Delta n$ within the interval $\Delta (\ve^2)$ is
\be\n{Dn}
\Delta n={1\over 2\hbar V_1}\Delta (\ve^2)\, .
\ee
The coefficient $1/ (2\hbar V_1)$ does not depend on $n$. In this sense, the corresponding distribution of energy levels, $\Delta n/\Delta (\ve^2)$, is equidistant.

\subsection{Parabolic potential: Nonlocal theory}

\subsubsection{Phase trajectories}

\begin{figure}[tbp]
\includegraphics[width=9.0cm]{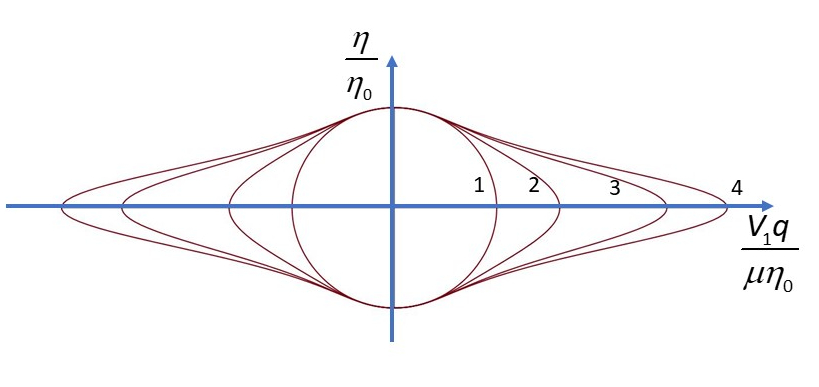}
  \caption{
  Phase trajectories  for the form factor $f=\exp(\hz^2)$. Different curves correspond to the different values of $\hz_0$ parameter: $0.14$ for line 1; $0.7$ for line 2;
    $1.0$ for line 3 and $1.1$ for line 4.
\label{P2}}
\end{figure}

In order to discuss the effect of the nonlocality on the distribution of the energy levels for the trapped nonlocal field we consider the same parabolic potential (\ref{VC}) as in the previous subsection, but modify the kinetic part of the effective Hamiltonian.
We use dimensionless units defined in the footnote~\ref{mu} and denote
\be
\hq=V_1 q/\mu\hh \hz_0=\he_0^2={1\over \mu^2}(\ve^2-m^2-p_{\perp}^2)\, .
\ee
Then the constraint  (\ref{CE}) gives
\be\n{hq}
\hq=\pm  \hF(\hz)^{1/2}\, .
\ee
Let us denote
\be \n{PT}
\he=\he_0 \sin\phi\hh \hz=\hz_0 \cos^2\phi \, .
 \ee
Then relations (\ref{hq}) and (\ref{PT}) allow one  to write the equation for phase trajectories in the parametric form
$(q(\phi),\eta(\phi))$.
The phase trajectories for the special choice of the form factor (\ref{cfz})  are shown at figure~\ref{P2}.

\subsubsection{Energy levels}

Using dimensionless variables the action integral (\ref{JS}), which enters the Bohr-Sommerfeld relation (\ref{BS}) for the energy levels, can be written as follows
\be \n{Jc}
J={2\mu^2\over V_1}\int_0^{\hz_0}\, {d\hz \sqrt{z}\over \sqrt{\hz_0-\hz} }f^{1/2}(\hz)\, .
\ee
We use here the expression for $\hF$ in terms of the form factor $f$, $\hF=\hz f(\hz)$.
After change of variables, $\hz=\hz_0 \xi$, relation (\ref{Jc}) takes the form
\be
J=\sigma(\hz_0) J_0\, .
\ee
Here $J_0$ is the value of the action integral for the local theory (\ref{JL}) and the factor $\sigma(\hz_0)$ is
\be \n{iii}
\sigma(\hz_0)={2\over \pi} \int_0^1 {d\xi \sqrt{\xi}\over \sqrt{1-\xi}} f^{1/2}(\hz_0 \xi)\, .
\ee

For a fixed value of $\eta_0$ and $\mu\to \infty$ the parameter $\hz_0\to 0$. Since $f(0)=1$  the integral (\ref{iii}) simplifies. It can be calculated analytically and the result is $\sigma(0)=1$.
Hence in this limit $J$  coincides with  $J_0$.  This is not surprising since in the limit $\mu\to\infty$ the theory becomes local. For a numerical calculation of $\sigma(\hz_0)$ it is convenient to put $\xi=\cos^2\phi$ and rewrite (\ref{iii}) in the form
\be  \n{iiii}
\sigma(\hz_0)={4\over \pi} \int_0^{\pi/2} d\phi\, \cos^2\phi  f^{1/2}(\hz_0 \cos^2\phi)\, .
\ee

\begin{figure}[hbt]
    \centering
    \vspace{10pt}
    \includegraphics[width=0.46\textwidth]{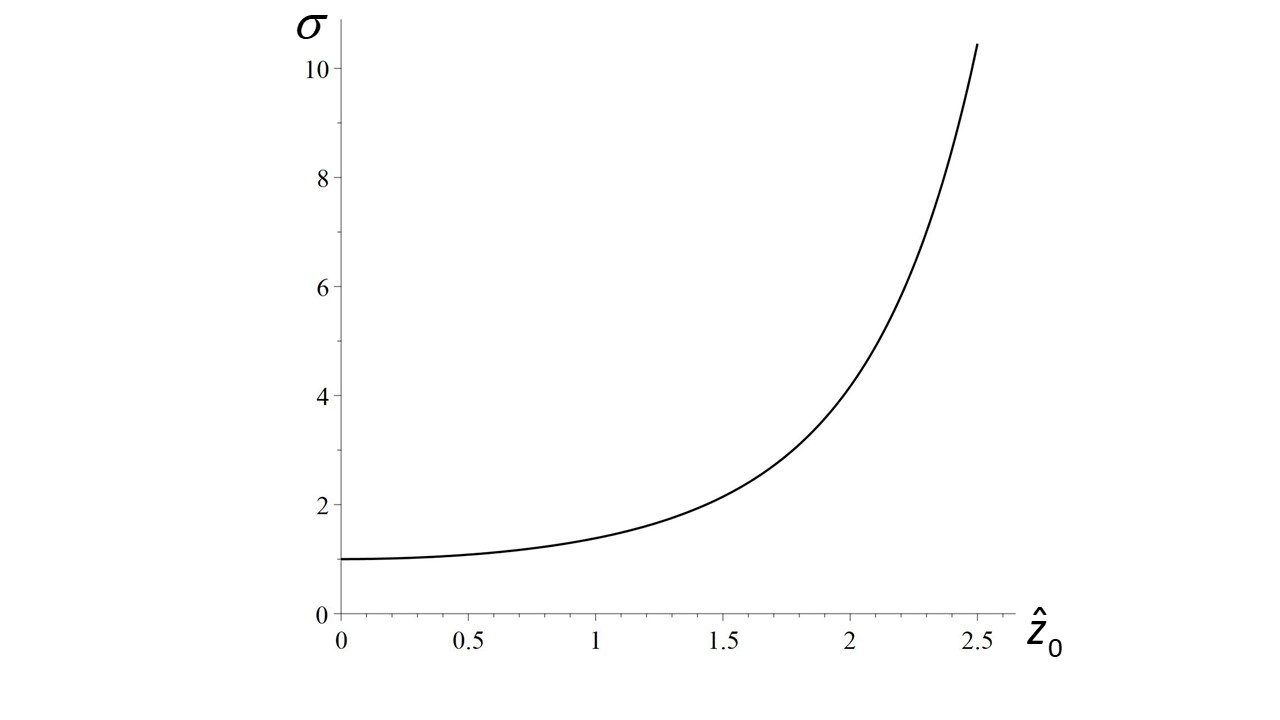}
    \caption{Function $\sigma(\hz_0)$ for the form factor $f(\hz)=\exp(\hz^2)$.}
    \label{P2a}
\end{figure}

\begin{figure}[hbt]
    \centering
    \vspace{10pt}
    \includegraphics[width=0.42\textwidth]{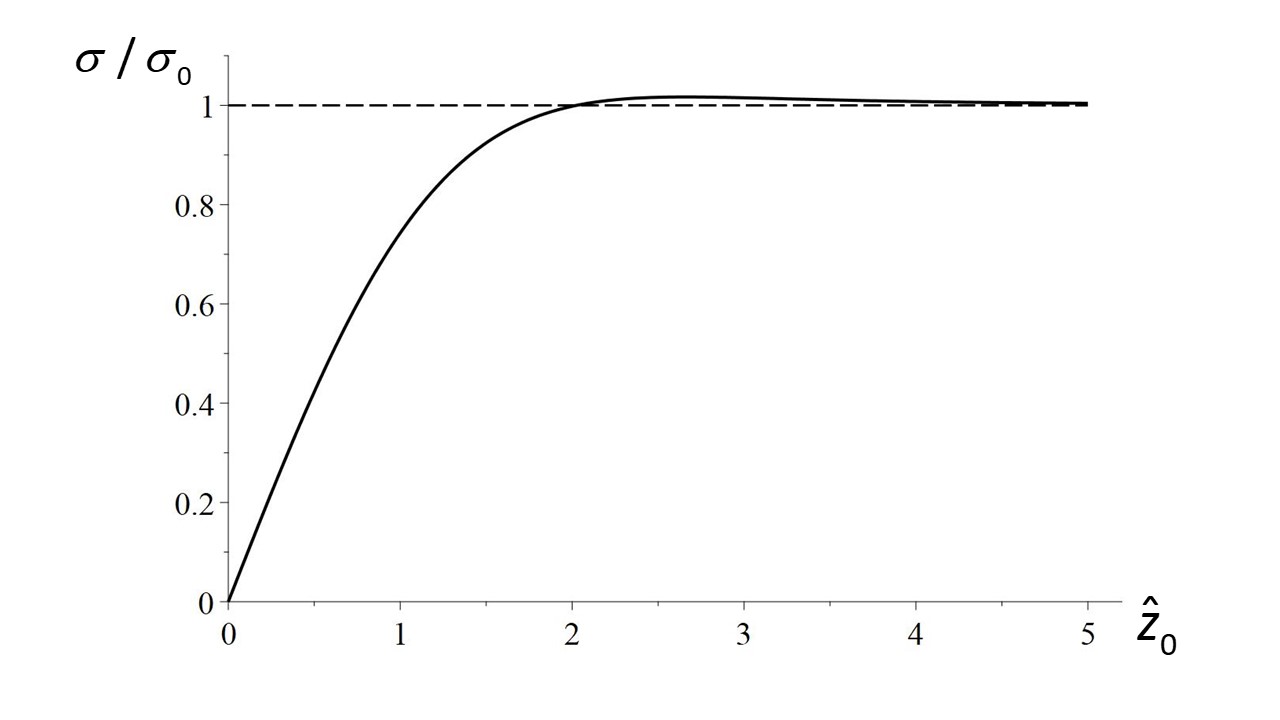}
    \caption{Ratio $\sigma/\sigma_0$ as a function of $\hz_0$.}
    \label{P3}
\end{figure}

The function $\sigma(\hz_0)$ for the choice of the form factor $f(\hz)=\exp(\hz^2)$ is shown at figure~\ref{P2a}. For this form factor
it is possible to show that  for large $\hz_0$ the function $\sigma(\hz_0)$ has the follows asymptotic form
\be\n{ss}
\sigma(\hz_0)\sim  {1.13\over  \hz_0}\exp(\hz_0^2/2)\, .
\ee
Figure~\ref{P3} which plots the ratio $ \sigma(\hz_0)/\sigma_0(\hz_0)$ demonstrates this property.

The parameter $\hz_0$ which enters $J$ depends on the energy $\ve$. The condition of quantization
\be\n{quan}
J=\pi\hbar (2n+1)
\ee
allows one to find the corresponding quasiclassical energy levels $\ve_n$. Since for small $\hz_0$ the coefficient $\sigma(\hz_0)$ is close to 1, the number of the energy levels $n$ in the interval $\Delta \ve^2$ has the form (\ref{Dn}). However for large $n$ this distribution of the number of levels is quite different. For the form factor $f(\hz)=\exp(\hz^2)$ one can find it by using the asymptotic form of the coefficient $\sigma(\hz_0)$
\be
J=1.13 {\pi \mu^2\over V_1} \exp(\hz_0^2/2)\, .
\ee
The condition of quantization (\ref{quan}) gives
\be
\exp(\hz_0^2/2)=0.88 {\hbar V_1\over \mu^2}(2n+1)\, .
\ee
For  $n\gg 1$ one gets
\be
\hz_0^2 \approx 2\ln n\, .
\ee
Since $\hz_0=\eta_0^2/\mu^2$, one finds
\be
\ve^2\approx m^2+p_{\perp}^2 +\sqrt{2}\mu^2 \sqrt{\ln n}\, .
\ee
This relation implies
\be \n{nnn}
\Delta n\approx {\sqrt{2}\over \mu^2} n\sqrt{\ln n} \, \Delta(\ve^2)\, .
\ee
The obtained result means that the equidistance of the energy level distribution (\ref{Dn}) of the local theory is broken by nonlocality. In the limit of large $n$ and given interval  $\Delta(\ve^2)$ the number of the corresponding levels $\Delta n$ grows with $n$ as $n\sqrt{\ln n}$.

\section{Barrier penetration}

\subsection{WKB approximation for under-barrier "motion"}
\begin{figure}[hbt]
    \centering
    \vspace{10pt}
    \includegraphics[width=0.40\textwidth]{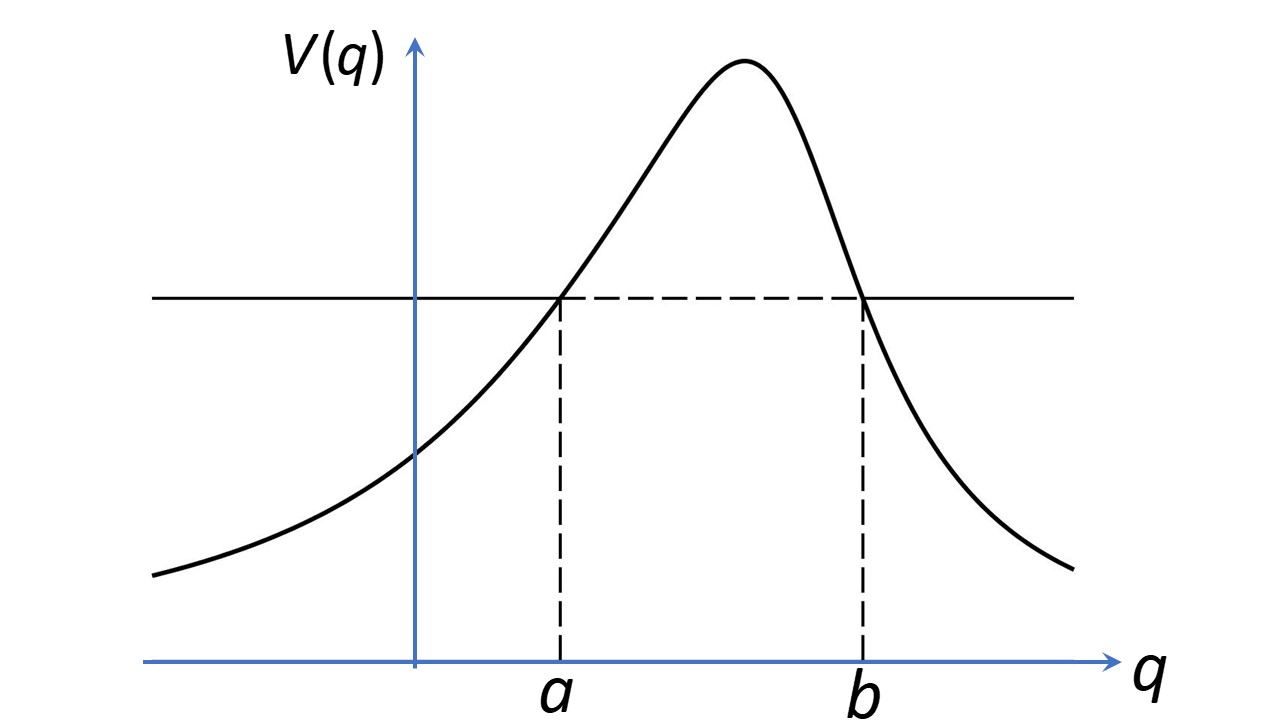}
    \caption{Potential $V(q)$.}
    \label{P6}
\end{figure}

 Let us consider a potential which has the form shown at figure~\ref{P6} and assume that the effective energy of the "particle" is such that its trajectory has a turning point at $q=a$. At this point one has
 \be
 \cF(\hz_0)=\hV(a)\hh \hz_0=\he_0^2\, .
 \ee
Let us write the constraint equation (\ref{CE}) in the form
\be \n{ub}
\cF(\hz_0)-\cF(\hz)=\hV(a)-\hV(q)\, .
\ee
In the domain to the left from the turning point $a$ one has $\hV(q)<\hV(a)$. We assume that for real $\hz$ the  function $\cF(\hz)$ is monotonically growing. Then one has $\hz<\hz_0$. In the region $q>a$ where $\hV(q)>\hV(a)$ one should have $\hz>\hz_0$. This condition implies that $\eta^2<0$ which is impossible for a real value of $\eta$.

It is well known that in the local theory a quasiclassical solution for a wave function in the classically forbidden  domain can be obtained by considering complex trajectories. Namely, one should change $\eta\to i\eta$. The same  trick does work for the non-local theories considered in this paper. After the change $\hz\to \he_0^2+\he^2$
the value of $\hz$ can be larger than $\hz_0$ for real $\he$ and the equation (\ref{ub}) has a solution in the region where $q>a$, that is in the classically forbidden domain.

A quasiclassical solution in the underbarrier domain can be obtained by the change $\eta\to i\eta$ in (\ref{PH}) and it is of the form
\ba \n{UPH}
&&\Phi(q)={C_-\over \sqrt{J}} \exp[- {1\over \hbar} {\cal S}(q)]+{C_+\over \sqrt{J}} \exp[ {1\over \hbar} {\cal S}(q)]\, ,\\
&&{\cal S}(q)= \int_{q_0}^{q} \eta dq=\int_{0}^{\tau} \eta(\tau) \dot{q}(\tau) d\tau \, .
\ea
Here
\be \n{Jb}
J=\mu \cF' \sqrt{\hz-\hz_0}\hh \hz_0=\he_0^2\, .
\ee
For real coefficients $C_{\pm}$  the WKB solution (\ref{UPH}) is real.

\subsection{Nonlocal field in a linear potential}

For a linear potential
\be
V(q)=U_0+U_1 (q-a)\, ,
\ee
one-dimensional nonlocal field equation (\ref{O1}) can be solved exactly. For this purpose we denote $q=a+x$ and write $\Phi(x)$ in the form
\be\n{Px}
\Phi(x)=\int_{-\infty}^{\infty}  d\eta\, e^{i\eta x/\hbar}\,  \tilde{\Phi}(\eta)\, .
\ee
Using the inverse Fourier transform one gets
\be
\tilde{\Phi}(\eta)={1\over 2\pi\hbar}\int_{-\infty}^{\infty} dx\,  e^{-i\eta x/\hbar} \, \Phi(x)\, .
\ee
The field equation (\ref{O1}) implies the following differential equation for $\tilde{\Phi}(\eta)$
\be
\left[ U_1{\hbar\over i} {\partial\over \partial \eta} +U_0+\mu^2\cF(\hz)\right]\tilde{\Phi}(\eta)=0\,  .
\ee
A solution of this equation is
\ba
&&\tilde{\Phi}(\eta)=C\exp\left[-{i\over \hbar U_1}{\cal S}(\eta)\right]\, ,\\
&& {\cal S}(\eta)=U_0\eta+\mu^2\int_0^{\eta} d\eta \cF(\hz)\, .
\ea
Substitution of this expression into (\ref{Px}) gives
\be
\Phi(x)=C\int_{-\infty}^{\infty}  d\eta \ \exp\left[{i\over \hbar} \left(\eta x-{1\over U_1}{\cal S}(\eta)\right)\right]\, .
\ee
This is a desired solution of the nonlocal field equation (\ref{O1})

\subsection{Connection of WKB solutions at a turning point}

Consider a potential of the form shown at figure~\ref{P6} and denote by $a$ a coordinate of a left turning point. One has $V_{,q}(a)>0$. To the left of $a$ and at some distance from it one can use WKB solution (\ref{PH}).
For the real field $\Phi(q)$ we denote
\be
C_1={1\over 2}C \exp(-i\psi)\hh
C_2={1\over 2}C \exp(i\psi)\,  ,
\ee
where $C$ and $\psi$ is real and  write the WKB solution (\ref{PH}) in the form
\ba  \n{LEFT}
&&\Phi(q)={C\over \sqrt{J}} \cos\left( {\cal S}/\hbar -\psi\right)\, ,\\
&&{\cal S}=\int_{a}^{q} \eta(q) dq\hh J=\mu \cF' \sqrt{\hz_0-\hz}\, .
\ea
To the right of the turning point $a$ and at some distance from it we use a decreasing in the under-barrier domain
WKB solution
\ba  \n{RIGHT}
&&\Phi(q)={C_-\over \sqrt{J}} \exp(- {\cal S}/\hbar)\, ,\\
&&{\cal S}= \int_{a}^{q} \eta(q) dq\hh J=\mu \cF' \sqrt{\hz-\hz_0}\, .
\ea

At the turning point the WKB approximation is not valid. In order to establish a relation between (\ref{LEFT}) and (\ref{RIGHT}) solutions one can use the following trick \cite{Landau:1965} which can be easily adapted to our problem. For this purpose we consider formally $\Phi(q)$ as a function of a complex variable $q$ and find it along a path on the complex plane of $q$ connecting  positive and negative values of $q-a$. We choose this path to be a half of the circle  of constant radius $\rho$ with the center at $q=a$ and take $\rho$ so that the conditions of the validity of the WKB approximation are satisfied along the path. Let us denote
\be \n{qa}
q-a=\rho e^{\pm \phi}\, .
\ee
We choose a sign $+$ for a semi-circle in the upper half of the complex $q$-plane, and $-$ for a semi-circle in the lower half of the complex $q$-plane. The angle $\phi$ changes from $0$ to $\pi$.
At the turning point $\hz=\hz_0=\he_0^2$ and $q=a$. We assume that $\rho$ is small so that $|\hz-\hz_0|$ and $|q-a|$ are also small and use the following linearized form of the constraint equation
\be\n{qz}
\cF'_0 (\hz-\hz_0)=\hV_{,q}(a) (q-a)\, ,\ \ \left. \cF'_0=(d\cF/d\hz)\right|_{\hz_0}\, .
\ee
Let us denote $B=\sqrt{\hV_{,q}(a)/\cF'_0}$. One also has $\hz-\hz_0=\he^2$.
Then (\ref{qa}) and (\ref{qz}) imply
\be
\eta=\mu B \rho^{1/2} e^{\pm i\phi/2}\, .
\ee
For the analytically continued WKB solution (\ref{RIGHT}) we get
\ba
&&{\cal S}^{\pm}={2\over 3}\mu B \rho^{3/2} \exp(\pm 3i\phi/2)\, ,\\
&&J^{\pm}=\mu B \cF'_0 \rho^{1/2} \exp(\pm i\phi/2)\, .
\ea
Here $\pm$ stands for the quantities calculated along a semi-circle in the upper ($+$) and lower ($-$) half of the $q$-plane. Using these relations at $\phi=\pi$ where $a-q=\rho$ one obtains
\be \n{CF}
C={1\over 2} C_-\hh \psi=-\pi/4\, .
\ee
This establishes a relation between the WKB solution (\ref{RIGHT}) in the domain to the right of the turning point and the WKB solution (\ref{LEFT}) to the left from it. This gives a required connection formula. It should be emphasized that for the operator $F(\Box)$ with the properties specified in section~II this connection relation is the same as for a standard Schr\"{o}dinger equation.

We assume now that the potential has two turning points $a<b$ and the domain between these points is forbidden for the classical motion.
We suppose that in the region $q<a$ the field $\Phi(q)$ has the form (\ref{LEFT}) and its amplitude is $C$. Then in the region $q>b$ it has the same form with a different amplitude $\tilde{C}$. Similarly to the local case the ratio $|\tilde{C}/C|^2$ can be found by
 using the connection formula (\ref{CF}) twice \cite{Landau:1965}
 \ba
 &&R=\left({\tilde{C}\over C}\right)^2 =\exp(-{2\over \hbar} P)\, ,\\
 &&P=\mu \int_a^b  \sqrt{\hz-\hz_0}\  dq\, .
 \ea
Here $\hz_0=\he_0^2$ and $\hz=\hz_0^2+\he^2$. In the expression for $P$ the quantities $\hz$  and $q$ satisfy the constraint (\ref{CE}).
The quantity $R$ gives a probability of the barrier penetration. It should be emphasized that these relations are obtained in the WKB approximation which is valid when $R\ll 1$.

\subsection{Barrier penetration}

\subsubsection{Inverse parabolic potential: Local theory}

\begin{figure}[hbt]
    \centering
    \vspace{10pt}
    \includegraphics[width=0.40\textwidth]{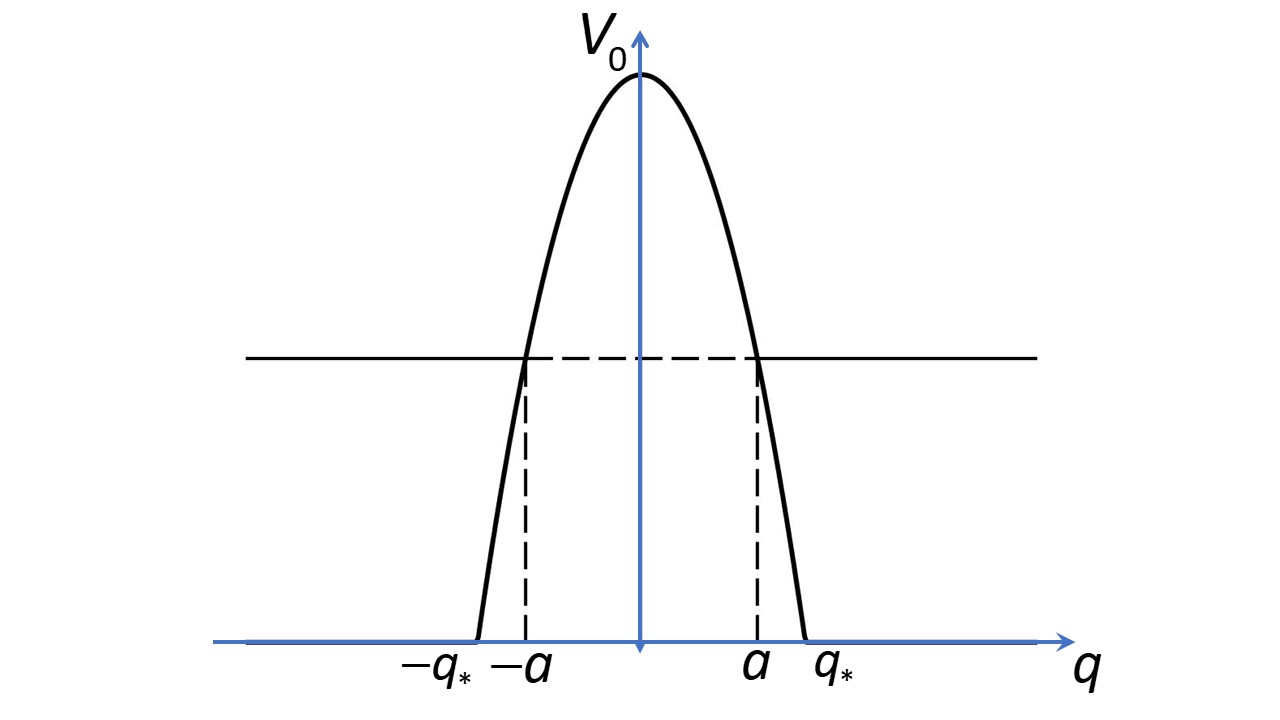}
    \caption{The inverse parabolic potential}
    \label{P4}
\end{figure}

For the local theory the effective Hamiltonian  is
\be
H(q,\eta)={1\over 2}[-(\eta_0^2 -\eta^2)+V(q)]\, ,
\ee
where $\eta_0^2=\ve^2-p_{\perp}^2-m^2$.
Let  the potential $V(q)$ be of the form
\be\n{INV}
V(q)=
\begin{cases}
V_0-V_1^2 q^2\, ,& \text{if\ \  \ } |q|\le q_*\, ,\\
 0 \, , &\text{if\ \  \ } |q|> q_*\,  .
\end{cases}
\ee
where  $V_0>0$ and $q_*={\sqrt{V_0}/ V_1}$. This potential is shown at figure~\ref{P4}. It is called an inverse parabolic potential.

We assume that $\eta_0^2<V_0$ so that there exit two turning points at $q=\pm a$, where
\be \n{aa}
a={\sqrt{V_0-\eta_0^2}\over V_1}\, .
\ee
For the motion in the under-barrier domain we put $\eta=i\tilde{\eta}$ where
\be
\tilde{\eta}=\pm \sqrt{V-\eta_0^2}=\pm V_1 \sqrt{a^2-q^2}\, .
\ee

We denote by $P_0$ the following integral
\ba
P_0&=&\int_{-a}^a |\tilde{\eta}|dq\nonumber\\
&=&V_1 \int_{-a}^a dq \sqrt{a^2-q^2}={1\over 2}\pi a^2 V_1\, .
\ea
Using relation (\ref{aa}) we can also write $P_0$ in the form
\be\n{P00}
P_0={\pi(V_0-\eta_0^2)\over  2 V_1}\, .
\ee
Let us emphasize that $P_0$ depends only on the difference $V_0-\eta_0^2$ but it does not depend on the value of $\eta_0$ itself. In other words, the penetration probability is sensitive only to the fact how close the effective energy $\eta_0^2$ is to the top of the potential $V_0$. Let us also remind that the adopted WKB approximation is valid only when the dimensionless quantity $P_0$ is large, that is when the probability of the penetration $R_0=\exp(-2P_0/\hbar)$  is small (see e,g, \cite{Landau:1965}).

\subsubsection{Inverse parabolic potential: Nonlocal theory}

Let us assume again that the potential $V(q)$ has the form (\ref{INV}). We denote by $\hz_0$ and $\hz_1$ solutions of the following equations
\be
\cF(\hz_0)=\hV_0 -\hV_1^2 a^2\hh
\cF(\hz_1)=\hV_0\, .
\ee
Then the WKB probability of the barrier penetration is
\ba
&&R=\exp\left(-{2\over \hbar}P\right)\, ,\\
&&
P={\mu \over \hV_1} \int_{\hz_0}^{\hz_1} {d\hz \over \sqrt{\hz-\hz_0}}\sqrt{\cF(\hz_1)-\cF(\hz)}\, .\n{PPP}
\ea

Let us calculate the value of $P$ for the case when $\cF$ is of the form (\ref{cfz}). When both parameters $V_0$ and $\eta_0^2$ are much smaller than $\mu^2$, the function $X(\hz)\approx\hz$ and the integral in (\ref{PPP}) can be easily taken with the following result
\be
P={\pi \mu (\hz_1-\hz_0)\over 2 \hV_1}={\pi(V_0-\eta_0^2)\over  2 V_1}\, .
\ee
In this regime the effects of the nonlocality are not important and we reproduce the expression for $P_0$ obtained in the local case (see equation (\ref{P00})).

\begin{figure}[hbt]
    \centering
    \vspace{10pt}
    \includegraphics[width=0.40\textwidth]{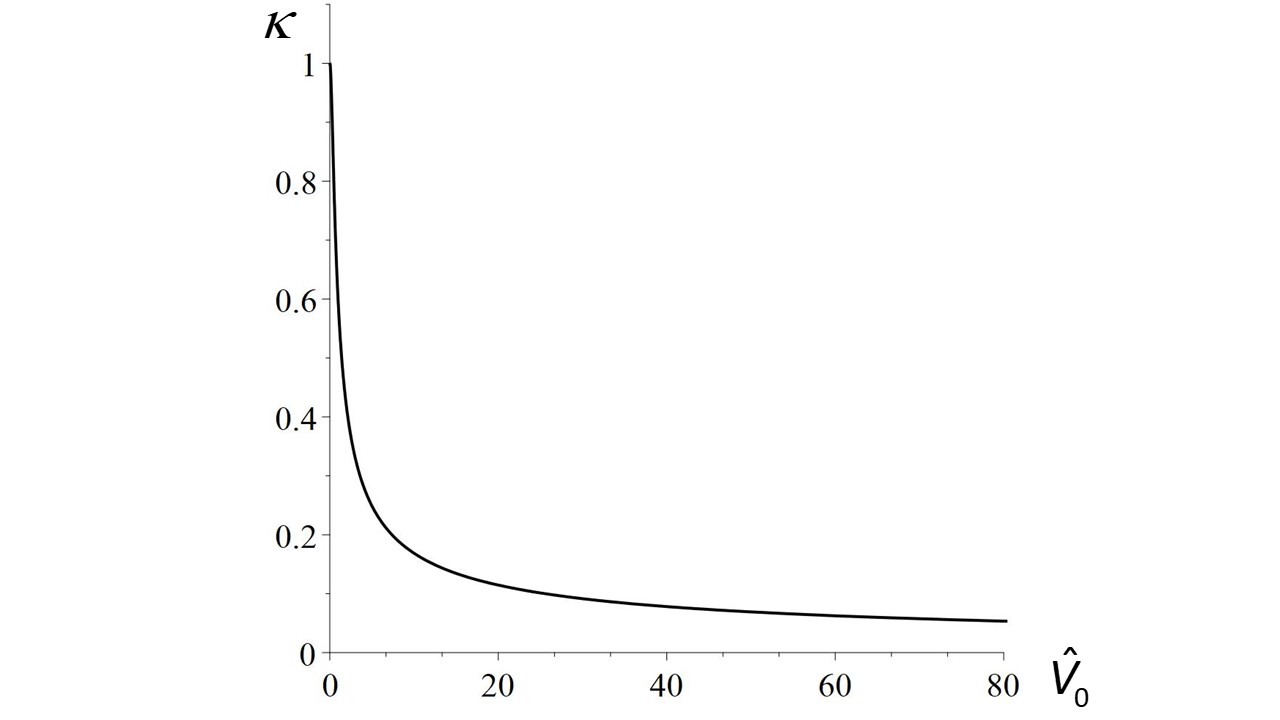}
    \caption{Parameter $\kappa$ as a function of $\hV_0$.}
    \label{P5}
\end{figure}

The effects of the nonlocality play an important role in the different regime, when $\hV_0\gg 1$ and $\he_0^2\gg1$.
To demonstrate this we consider again the theory with the form factor $f(\hz)=\exp(\hz^2)$.
In order to compare the expression for $P$ with $P_0$ we fix as earlier the value of the difference $V_0-\eta_0^2$,  but now we assume that $\hV_0\gg1$. Then one has $\hz_1-\hz_0\ll 1$.
In the appendix~B it is shown that one can write the following approximate expression for $P$ in this case
\be\n{kkk}
P=\kappa(\hV_0) P_0\hh \kappa(\hV_0)={1\over \sqrt{X'_1}}.
\ee
Here $\left. X'_1=dX/dx\right|_{x=\hz_1}$ and $X(\hz_1)=\hV_0$.
Using (\ref{der}) one can write
\ba
&&X'_1(\hz_1)=(1+2\hz_1^2) \exp(\hz_1^2)\, ,\\
&&\hV_0=\hz_1^2  \exp(\hz_1^2)\, .
\ea
These relations give a parametric representation of $X'_1(\hz_1)$ as a function of $\hV_0$. The  plot of the function
$\kappa(\hV_0)$ is shown at figure~\ref{P5}. It shows that this function is equal to 1 at $\hV_0=0$ and it is less than 1 and monotonically decreases for positive $\hV_0$.

Since $R=\exp(-2P/\hbar)$ the relation (\ref{kkk}) can be rewritten as follows
\be
R=(R_0)^{\kappa(\hV_0)}\, .
\ee
Since $R_0<1$  the value of $R$ is always greater that $R_0$. This means that in the nonlocal theory the probability of the barrier penetration is larger that in the local case. In other words nonlocality makes the barrier more transparent in the high-energy regime.

\section{Discission}

Let us discuss the obtained results. In this paper we study quasiclassical solutions of the nonlocal massive scalar field equations. The kinetic part of the action for such a theory contains a function of the $\Box$-operator, which is chosen so that the theory does not contain new unphysical degrees of freedom. In section~II we formulated conditions imposed on the form factor of such a theory. We also described a special model which is used later in the paper for the illustration. In section~III we described an ansatz used in the WKB approximation for a solution of the nonlocal field equation. In the leading order of the $1/\hbar$-expansion one obtains the equation for the eikonal function $S(x)$ which is of the form $H(x,\nabla S)=0$. This is a first order partial differential equation which has the form of the Hamilton-Jacobi equation for the Hamiltonian $H(x,p)$, obtained by substitution $\nabla S=p$. To calculate the eikonal function $S(x)$ it is sufficixent to construct the Lagrangian submanifold in the phase space $(x,p)$ by using the initial conditions for $S$. Such a Lagrangian submanifold is formed by phase trajectories which are solutions of the Hamiltonian equations of motion. Since the system is conservative this submanifold belongs to the subspace $H(x,p)=0$.
Let us emphasize that this procedure is very similar to the standard one adopted for the quantum mechanics and local field theory. When the field operator is of the second order the corresponding effective Hamiltonian $H(x,p)$ is a quadratic polynomial in momentum $p$. For the nonlocal theory  $H(x,p)$ is a non-polynomial function of $p$ and this is a main difference with the local case.
As a result the construction of the WKB solutions becomes technically  more complicated.

WKB solutions presented in section~III are valid for a wide class of nonlocal scalar field theories with an
arbitrary external potential $V(x)$ in a flat spacetime with any number of spatial dimensions. These relations are greatly simplified for the case when the potential $V$ depends only  on  one spatial Cartesian coordinate. For this case
the phase space is two-dimensional and the Hamilton's equations are completely integrable. The corresponding WKB solutions are described in section~IV. In section~V we used the WKB approximation for the calculation of the energy levels of the nonlocal field trapped by a one-dimensional potential. To illustrate general formulas we considered the case when the form factor of the nonlocality is of the special form (\ref{cfz}) and the potential is parabolic. The energy levels $\ve_n$ of the field depend on a discrete integer parameter $n$.
For the local case  the number $\Delta n$ of the levels in the interval $\Delta (\ve^2)$
is $\Delta n=\Delta (\ve^2)/(2\hbar V_1)$ and it does not depend on the value of $n$. In this sense the levels are equidistant. In the nonlocal case for large $n$ this dependence is different
\be
\Delta n\approx {\sqrt{2}\over \mu^2} n\sqrt{\ln n} \, \Delta(\ve^2)\, .
\ee
This means that for a fixed value of $\hbar V_1/\mu^2$ and very high energy there are much more energy levels 
in a given interval of $\Delta(\ve^2)$  than in the local case.

Finally, we discussed the under-barrier motion and obtained an expression for the probability of the barrier penetration.
This probability has the form $ \sim \exp(-2P/\hbar)$. We calculated the factor $P$ for the the special form of the nonlocality  (\ref{cfz}) in the presence of  the inverse parabolic  potential. For the local theory this factor is
$P_0=\pi (V_0-\eta_0^2)/(2 V_1)$. In other words, it depends on how close is the energy of the state $\eta_0^2$ to the top of the potential $V_0$, but it does not depend on the height $V_0$ of the potential itself. The situation in the nonlocal case is quite different. Namely, one has $P=\kappa P_0$, where the factor $\kappa$ depends on the height $V_0$ of the potential. This factor is always less than 1 and it decreases when $V_0$ grows. This means that the nonlocality "helps" the field to penetrate the potential barrier.

The examples considered in this paper are rather simple. It would be interesting to  apply the developed in the paper method for study  the effects of the nonlocality in the early cosmology and in the black holes interior.

\appendix

\section{WKB approximation}
\n{AA}
We collected here useful information concerning construction of solutions of the nonlocal field equations in the WKB approximation. Further details and required mathematical foundations can be found in the book \cite{maslov1981semi}.

Let $(x^1,\ldots,x^N)$ be coordinates of a point $x$ in $R^N$. We denote by $\alpha$ a set of $N$ non-negative integer numbers $\alpha=(\alpha_1,\ldots,\alpha_N)$.
We consider the following operator in this space
\be \n{OO}
\hat{\cal O}=\sum_{\alpha} a_{\alpha} \left( {\hbar \over  i}{\partial\over \partial x^1}\right)^{\alpha_1}\ldots
\left( {\hbar \over  i}{\partial\over \partial x^N}\right)^{\alpha_N}\, .
\ee
The sum is taken over all different combinations of $\alpha$. We do not assume that the number of terms  in the sum is finite, so that this expression can be used as a formal series in derivatives representation of nonlocal operators.

Let us make a substitution
\be
 {\hbar \over  i}{\partial\over \partial x^{\mu}}\to p_{\mu}
\ee
in the expression (\ref{OO}). Then we obtain a function of $2N$ variables which we denote by $H=H(x,p)$
\be \n{HH}
H=H(x,p)=\sum_{\alpha} a_{\alpha} p_1^{\alpha_1}\ldots p_{N}^{\alpha_N}\, .
\ee
Here $p=(p_1,\ldots,p_N)$. This object is called a symbol of the operator $\hat{\cal O}$.

We are looking for a solution $\varphi$ of the equation
\be \n{DF}
\hat{\cal O}\varphi=0
\ee
in the form of the following formal series
\be \n{vv}
\varphi(x)=\exp\left({i S(x)\over \hbar}\right)\sum_{j=0}^{\infty} \left({\hbar\over i}\right)^j  u_j(x)\, .
\ee
Usually it is sufficient to consider only first few terms in the series. This gives what is known as the WKB approximation for the solution $\varphi$. To obtain the functions $S(x)$ and $u_j(x)$ that enter (\ref{vv}) the following procedure is used (see e.g. \cite{Arnold:1989,maslov1981semi} and references therein).

If one substitute expression (\ref{vv}) into the equation (\ref{DF}) then one finds that in the leading order $j=0$ the following condition is to be satisfied
\be
H\left(x,{\partial S\over \partial x}\right) u_0=0\, ,
\ee
where
\be
{\partial S\over \partial x}=\left( {\partial S\over \partial x^1}\ldots, {\partial S\over \partial x^N}\right)\, .
\ee
For a nonvanishing function $u_0(x)$ the following Hamilton-Jacobi equation is valid
\be\n{HJ}
H\left(x,{\partial S\over \partial x}\right)=0\, .
\ee

Let us consider a $(2N)$-dimension space $\Gamma^{2N}$ with coordinates $(x,p)=(x^1,\ldots,x^N,p_1,\ldots,P_N)$ and denote by $\Omega$ a symplectic two-form in it
\be
\Omega=\sum_{\mu=1}^N dp_{\mu}\wedge dx^{\mu}\, .
\ee
An $N$-dimensional surface $\Lambda^N$ in $\Gamma^{2N}$ is called a Lagrangian submanifold if the $\Omega$  vanishes on $\Lambda^N$. If a Lagrangian submanifold can be uniquely projected onto the $x$-space, then there exists such a function $S(x)$ that   $p_{\mu}=\partial S/\partial_{\mu}$ on $\Lambda^N$.

The Hamiltonian $H(x,p)$ given by the relation (\ref{HH}) implies the following Hamilton's equations
\be \n{HE}
{dx^{\mu}\over d\tau}={\partial H(x,p)\over \partial p_{\mu}}\hh
{dp_{\mu}\over d\tau}=-{\partial H(x,p)\over \partial x^{\mu}}\, .
\ee
These equations determine the Hamiltonian flow in the phase space $\Gamma^{2N}$ preserving the symplectic form $\Omega$ and  the Hamiltonian itself is an integral of motion.

Let $\Sigma^{N-1}$ be a $(N-1)$-dimensional surface in the coordinate space. Its  embedding is defined by the equations
\be
x^{\mu}=f^{\mu}(y^i)\hh i=1,\ldots,N-1\, .
\ee
One can consider $y^i$ as coordinates on $\Sigma^{N-1}$. We associate with $\Sigma^{N-1}$ a $(2N-2)$-dimensional phase space $\Gamma^{2N-2}$ and denote its canonical coordinates by $(y,v)=(y^1,\ldots,y^{N-1},v_1,\ldots,v_{N-1})$. The  symplectic 2-form $\omega$ in $\Gamma^{2N-2}$ is
\be
\omega=\sum_{i=1}^{N-1} dv_{i}\wedge dy^{i}\, .
\ee

Denote by $S^0=S^0(y)$ a function on $\Sigma^{N-1}$ and $\nabla S^0=(S^0_{,y^1},\ldots,S^0_{,y^{N-1}})$, then a $(N-1)$-dimensional surface $(y,\nabla S)$ of $\Gamma^{2N-2}$ is a Lagrangian submanifold $\Lambda^{N-1}$, so that $\omega|_{\Lambda^{N-1}}=0$.

Let us define a covector $p(y)=p_{\mu}(y)$ on $\Sigma^{N-1}$ by two conditions
\be\n{ICS}
p_{\mu}(y)dx^{\mu}=p_{\mu}(y){\partial f^{\mu}\over \partial y^i}=0\, , \
H(x(y),p(y))=0\, .
\ee
We assume that this system of $N$ equations has a solution which determines $p_{\mu}(y)|_{\Sigma^{N-1}}$.

Let us consider the following initial conditions for the Hamilton's equations (\ref{HE})
\be \n{IC}
x^{\mu}|_{\tau=0}=f^{\mu}(y^i)\hh p_{\mu}|_{\tau=0}=p_{\mu}(y)\, .
\ee
A family of solutions of the Hamilton's equations (\ref{HE}) with these initial conditions, $(x(\tau,y),p(\tau,y))$, forms a $N$-dimensional  surface in $\Gamma^{2N}$ which we denote by ${\cal B}$. It is possible to show that ${\cal B}$ is a Lagrangian submanifold in $\Gamma^{2N}$ and that $H(x,p)=0$ on it. This implies that there exists such a function $S(x)$ that
\be
p_{\mu}|_{{\cal B}}={\partial S\over \partial x^{\mu}}\, .
\ee
This function is a solution of the Hamilton-Jacobi equation (\ref{HJ}) satisfying the initial condition $S(x)|_{\Sigma^{N-1}}=S^{0}(y)$.

This function $S(x)$ can be found as follows. Denote a 1-form $\theta=\sum_{\mu} p_{\mu} dx^{\mu}$, then
$\Omega=d\theta$. Since the symplectic form $\Omega$ vanishes on ${\cal B}$, the integral
\be
\int p_{\mu} dx^{\mu}
\ee
calculated along a path on ${\cal B}$ between its two points does not depend on the choice of the path. The required solution $S(x)$ of the Hamilton-Jacobi equation satisfying the initial condition $S|_{\Sigma^{N-1}}=S(y)$ can be written in the form
\be
S(x)=S^0(y)+\int_{0}^{\tau} p_{\mu} dx^{\mu}\, .
\ee
Here $x^{\mu}=x^{\mu}(\tau,y)$ and $p_{\mu}=p_{\mu}(\tau,y)$.

Let us note that the map $x^{\mu}=x^{\mu}(\tau,y)$ may be considered as a transformation  between the coordinates $(\tau,y$ and $x^{\mu}$. Let us denote by $J$ a Jacobian of this transformation
\be\n{JJJ}
J(\tau,y)=\left| \det {\partial x(\tau,y)\over \partial (\tau,y)}\right| \, .
\ee
Then the leading term of the asymptotic solution (\ref{vv}) is \cite{maslov1981semi}
\ba
\varphi(x)&=&u_0(y) \sqrt{ J(0,y)\over J(\tau,y)} \exp \left[{ i\over \hbar}\left( S^0(y)+\int_{0}^{\tau} p_{\mu} dx^{\mu}\right)\right. \nonumber \\
&-& \left. {1\over 2}\int_0^{\tau} \sum_{\mu=1}^{N} {\partial^2 H(x,p)\over \partial x^{\mu}\partial p_{\mu} }d\tau \right]\, . \n{QC}
\ea
For $\tau=0$ the right-hand side of this expression reduces to $u_0(y)\exp \left[{ i\over \hbar} S^0(y)\right]$ and it should be found from the initial condition for the  function $\varphi(x)$ at $\Sigma$.

\section{Function \texorpdfstring{$X(x)=x \exp(x^2)$}{Lg}   and its properties}
\n{BB}

In this appendix we discuss some properties of the function
\be
X(x)=x \exp(x^2)\, .
\ee
The derivatives of this function are
\ba  \n{der}
&&X'=(1+2x^2)\exp(x^2)\, ,\nonumber\\
&&X''=6x(1+{2\over 3}x^2)\exp(x^2)\, ,\\
&&\alpha={1\over 2}{X''\over X'}=x\left(1+{2\over 2x^2+1} \right)\, .\nonumber
\ea
We denote by $Y$ an inverse function of $X$ such that the following relations are valid
\be
y=X(x)\hh x=Y(y)\, .
\ee
The inverse function can be written as follows
\be
Y(y)=y \exp\left(
-{1\over 2} W(2y^2)\, .
\right)
\ee
Here $W$ is Lambert $W$ function defined by the relation
\be
W(z) \exp(W( z))=z\, .
\ee
For a real positive argument  $W$  is a real single-valued function.

Consider a following integral
\be
I=\int_{x_0}^{x_1}{dx\over \sqrt{x-x_0}}\sqrt{X(x_1)-X(x)}\, .
\ee
Let us assume that $x_1-x_0$ is small. To obtain an approximate value of the integral $I$ in this case we use the Taylor expansion
\be\n{XX}
X(x)=X_1-X'_1(x_1-x)+{1\over 2}X''_1(x_1-x)^2+\ldots\, .
\ee
Here $X_1=X(x_1)$,  $X'_1=X'(x_1)$, $X''_1=X''(x_1)$.
Skipping the higher order terms denoted in (\ref{XX}) by dots one has
\be
I=\sqrt{X'_1}\int_{x_0}^{x_1}{dx\over \sqrt{x-x_0}}\sqrt{(x_1-x)-\alpha (x_1-x)^2}\, ,
\ee
where $\alpha={1\over 2}(X''_1/X'_1)$. This integral can be calculated exactly.
Let us denote $\beta=x_1-x_0$, then
for $\alpha \beta<1$ the answer is
\be \n{Ia}
I={2\sqrt{X'_1}\over 3\alpha}\left[
(2\alpha \beta -1) E(\sqrt{\alpha \beta})-(\alpha \beta-1)K(\sqrt{\alpha \beta})
\right]\, .\nonumber
\ee
Here $E$ and $K$ are complete elliptic integrals. Forthe small value of $\alpha \beta$ one has
\be
I={\pi \beta\over 2}\sqrt{X'_1} \left(1-{3\over 8}\alpha \beta\right)\, .
\ee
For $\alpha \beta\ll 1$ one can neglect the second term. Keeping the leading term one gets
\be  \n{Ib}
I={\pi \beta\over 2}\sqrt{X'_1} ={\pi\over 2} {X_1-X_0\over \sqrt{X'_1}}\, .
\ee
For small $x_1$ one has $X'_1\approx 1$ and
\be \n{Ic}
I\approx {\pi\over 2} (X_1-X_0)\, .
\ee

\section*{Acknowledgments}
The author thanks the Natural Sciences and Engineering Research Council of Canada and the Killam Trust for their financial support.


%

\end{document}